\documentclass{jfm}
\usepackage{hyperref}
\hypersetup{
	colorlinks=true,
	allcolors=blue}
\usepackage{graphicx}
\usepackage{epstopdf, epsfig}
\usepackage{amsmath}
\usepackage{commath}
\usepackage[section]{placeins}
\usepackage{float}

\usepackage{caption}
\usepackage{subcaption}
\usepackage{bm}
\usepackage{amssymb}
\usepackage{bigints}

\allowdisplaybreaks

\title{Inertial migration of a sphere in plane Couette flow}
\author{Prateek Anand\aff{1} and Ganesh Subramanian\aff{1}
  \corresp{\email{sganesh@jncasr.ac.in}}}

\affiliation{\aff{1}Engineering Mechanics Unit, Jawaharlal Nehru Centre for Advanced Scientific Research, Bengaluru-560064, India.}
\begin{document}
\maketitle

\begin{abstract}
We study the inertial migration of a torque-free neutrally buoyant sphere in wall-bounded plane Couette flow over a wide range of channel Reynolds numbers, $Re_c$, in the limit of small particle Reynolds number\,($Re_p\ll1$) and confinement ratio\,($\lambda\ll1$). Here, $Re_c = V_\text{wall}H/\nu$ where $H$ denotes the separation between the channel walls, $V_\text{wall}$ denotes the speed of the moving wall, and $\nu$ is the kinematic viscosity of the Newtonian suspending fluid; $\lambda = a/H$, $a$ being the sphere radius, with $Re_p=\lambda^2 Re_c$. The channel centerline is found to be the only (stable)\,equilibrium below a critical $Re_c\,(\approx 148)$, consistent with the predictions of earlier small-$Re_c$ analyses. A supercritical pitchfork bifurcation at the critical $Re_c$ creates a pair of stable off-center equilibria, symmetrically located with respect to the centerline, with the original centerline equilibrium simultaneously becoming unstable. The new equilibria migrate wallward with increasing $Re_c$. In contrast to the inference based on recent computations, the aforementioned bifurcation occurs for arbitrarily small $Re_p$ provided $\lambda$ is sufficiently small. An analogous bifurcation occurs in the two-dimensional scenario, that is, for a circular cylinder suspended freely in plane Couette flow, with the critical $Re_c$ being approximately $110$.
\end{abstract}


\section{Introduction}

One of the first observations of cross-stream migration of neutrally buoyant spherical particles, in an ambient shearing flow, were the experiments of \cite{segresilberberg1962a,segresilberberg1962b} involving pipe flow of a dilute suspension of such particles. For small values of the pipe Reynolds number, the spheres were found to accumulate at an intermediate annulus, about $0.6$ times the pipe radius from the centerline, creating a ``tubular pinch'' effect. Cross-stream migration of spheres in a uni-directional shearing flow is prohibited by the time-reversibility of the Stokes equations, and the observed migration is due to inertial lift forces, occurring only for non-zero particle Reynolds numbers. As described in more detail in \cite{anandJeffAvgd2022}, a number of theoretical and numerical studies have since attempted to explain these observations, the first one being that of \cite{holeal1974}, albeit for a shearing flow in a two-dimensional channel geometry. \cite{holeal1974} determined the inertial lift force on a sphere for both the plane Couette and plane Poiseuille velocity profiles, for $Re_c\ll1$, $Re_c$ being the channel Reynolds number. While the lift force in plane Poiseuille flow equalled zero at a location intermediate between the wall and the centerline, consistent with the observed intermediate equilibrium in the Segre-Silberberg experiments, that in plane Couette flow always pointed towards the channel centerline, rendering this location the only stable equilibrium. A subsequent more accurate calculation of the lift force profiles, again for $Re_c \ll 1$, was performed by \cite{vasseur1976} using a Green's function formulation developed earlier by \cite{coxbrenner1968}.

Later, \cite{feng1994} carried out finite element computations for a neutrally buoyant circular cylinder in plane Couette flow for $\lambda=0.125$ with $Re_p=0.625$ and $1.25$. Here, $\lambda$ is the confinement ratio, defined as the ratio of the cylinder radius to the inter-wall spacing, with $Re_p = \lambda^2 Re_c$ being the particle Reynolds number; $Re_c=40$ and $80$ for the cases above, with the authors still obtaining the channel centerline as the only stable equilibrium, consistent with the predictions of the aforementioned small-$Re_c$ analyses. Very recently, the motion of both a neutrally buoyant circular cylinder \citep{fox2020} and sphere \citep{fox2021}, again in plane Couette flow, was studied using lattice-Boltzmann simulations. The parameter ranges corresponding to $0.1\leq Re_p\leq50$ and $0.1\leq\lambda\leq0.2$\,($0.1\leq Re_p\leq50$ and $0.0625\leq\lambda\leq0.25$) were examined for the sphere (cylinder), with the associated maximum $Re_c$ significantly exceeding that in \cite{feng1994} above. The larger $Re_c$'s simulated led to the authors finding the emergence of stable off-centre equilibria via a pitchfork bifurcation. For a given $\lambda$, the particle migrated to the only (stable) equilibrium at the centerline below a critical $Re_p$, while above this critical value, a pitchfork bifurcation led to an unstable equilibrium at the center, with a pair of symmetrically located stable equilibria on either side of the centerline. While the simulations pointed to a decrease in the critical $Re_p$ with decreasing $\lambda$, owing to the limited range of parameters examined, it remained unclear as to whether the bifurcation was a signature of a finite $Re_p$\,(as implied, for instance, in the abstracts of the said articles), or if it correlated to a threshold level of inertia on scales of order the channel width. Herein, we show that the latter is the case. That is to say, the bifurcation threshold in plane Couette flow, for both a circular cylinder and a sphere, is shown to correspond to a critical $Re_c$ rather than $Re_p$, and this is accomplished via a point-particle formulation similar to the one used by \cite{schonberghinch1989} to study inertial migration in plane Poiseuille flow.

The paper is organized as follows. $\S$\ref{sec:Ch5formulation} discusses the governing equations and boundary conditions for a neutrally buoyant sphere suspended freely in plane Couette flow, for arbitrary $Re_c$, in the limit $Re_p \ll1$\,(implying $\lambda \ll 1$). In $\S$\ref{sec:Ch5Recsmall}, the inertial lift velocity is determined semi-analytically for $Re_c \ll 1$ using a reciprocal theorem formulation. Inertia acts as a regular perturbation in this limit, the lift velocity being $O(Re_p)$ and its calculation only requiring the knowledge of the velocity fields induced by a Stokeslet and a stresslet confined between (infinite)\,parallel plane boundaries. In $\S$\ref{sec:Ch5Reclarge}, the lift velocity is calculated for $Re_c\gtrsim O(1)$, with inertia now acting as a singular perturbation. Similar to \cite{schonberghinch1989}, this calculation involves obtaining coupled second-order ODEs for the partially Fourier transformed pressure and normal velocity fields, and then solving the associated boundary value problem numerically using a shooting method. The results for the equilibrium loci, in $\S$\ref{sec:Ch5Results}, reveal a supercritical pitchfork bifurcation at $Re_c\approx148$; the analogous calculation for a circular cylinder shows a bifurcation at $Re_c \approx 110$. We conclude in $\S$\ref{sec:Ch5conclusion} by discussing the effects of finite particle size on the bifurcation surface in $s_{eq}-Re_c-\lambda$ space, $s_{eq}$ denoting the (transverse)\,equilibrium position of the sphere within the channel.

\section{Formulation} \label{sec:Ch5formulation}
\begin{figure}
	\centering
	\includegraphics[width=\textwidth]{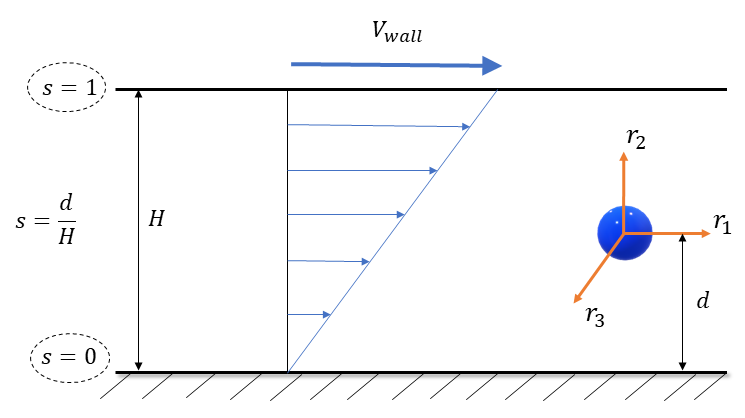}
	\caption{Schematic of a sphere suspended in wall-bounded plane Couette flow.}
	\label{fig:Ch5CouetteFlowGeometry}
\end{figure}
Figure \ref{fig:Ch5CouetteFlowGeometry} shows a schematic of a neutrally buoyant sphere suspended in a Newtonian fluid undergoing plane Couette flow between infinite plane boundaries separated by a distance $H$. Using the sphere radius $a$ and the velocity scale $\dot{\gamma}a$, $\dot{\gamma}=V_\text{wall}/H$ being the ambient shear rate, the equations governing the fluid motion, in non-dimensional form, are given by,
\begin{subequations}
	\begin{align}
	\nabla^2 \bm{u}-\bm{\nabla}p &=Re_p\,\bm{u\cdot\nabla u}, \label{eq:Ch5NS1}\\
	\bm{\nabla}\cdot\bm{u}&=0 \label{eq:Cont1},
	\end{align} \label{eqns:1}
\end{subequations}
where $Re_p=a^2\dot{\gamma}/\nu$ is the particle Reynolds number, $\nu$ being the kinematic viscosity of the suspending fluid. On account of the slow transverse motion arising from the inertial lift, the unsteady component of the acceleration in (\ref{eq:Ch5NS1}) has been neglected. For $\lambda \ll 1$, at leading order, the neutrally buoyant sphere translates with the velocity of the plane Couette flow at its center, while rotating with half the ambient vorticity. In a reference frame centered at the sphere, and translating with this velocity, $\bm{u}$ still satisfies (\ref{eq:Ch5NS1}) and (\ref{eq:Cont1}) with the following boundary conditions:
\begin{subequations}
	\begin{align}
	\bm{u}&=-\frac{1}{2}\bm{1}_3\wedge \bm{r} \text{ for } \bm{r}\in S_p,\\
	\bm{u}&\rightarrow r_2\bm{1}_1 \text{ for } r_1,r_3\rightarrow \infty, \\
	\bm{u}&=-\lambda^{-1}s\bm{1}_1\,\, \text{at}\,\,r_2=-s\lambda^{-1}\text{(lower wall)},\\
	\bm{u}&=-\lambda^{-1}(1-s)\bm{1}_1 \,\, \text{at}\,\, r_2=(1-s)\lambda^{-1} \text{ (upper wall)}.
	\end{align} \label{eq:Ch5BC1}
\end{subequations}
Here, $S_p$ denotes the surface of the sphere, and $s=d/H$ denotes its (non-dimensional)\,transverse location within the channel; see Figure \ref{fig:Ch5CouetteFlowGeometry}. In terms of the corresponding disturbance fields, $\bm{u}'=\bm{u}-\bm{u}^\infty$, $p'=p-p^\infty$\,($\bm{u}^\infty = r_2\bm{1}_1$, $p^\infty$ being a constant), the governing equations may be written as,
\begin{subequations}
	\begin{align}
	\nabla^2 \bm{u}'-\bm{\nabla} p' &=Re_p\left(\bm{u}'\cdot\bm{\nabla u}'+\bm{u}'\cdot\bm{\nabla u}^\infty+\bm{u}^\infty\cdot\bm{\nabla u}'\right),\\
	\bm{\nabla}\cdot \bm{u}'&=0,
	\end{align} \label{eq:Ch5NS2}
\end{subequations}
with boundary conditions (\ref{eq:Ch5BC1}a-d) taking the form:
\begin{subequations}
	\begin{align}
	\bm{u}'&=-\frac{1}{2}\bm{1}_3\wedge \bm{r}- r_2\bm{1}_1 \text{ for } \bm{r}\in S_p,\\
	\bm{u}'&\rightarrow 0 \text{ for } r_1,r_3\rightarrow \infty,\\
	\bm{u}'&=0\,\, \text{at}\,\, r_2=-s\lambda^{-1}, (1-s)\lambda^{-1}.
	\end{align} \label{eq:Ch5BC2}
\end{subequations}

It is well known that one requires a matched asymptotics expansions approach to solve the Navier Stokes equations for small but finite $Re_p$\citep{proudman_pearson_1957}. The matching of an inner expansion, valid in the neighbourhood of the particle, and an outer expansion, valid at distances of order the so-called inertial screening length, is in general necessary to calculate inertial corrections. For an ambient shearing flow, the screening length is $a Re_p^{-1/2}$, or alternatively, $H Re_c^{-1/2}$. The latter expression shows that the walls lie in the inner Stokesian region for $Re_c\ll1$ with fluid inertia being a regular perturbation\,($\S$\ref{sec:Ch5Recsmall} below). When $Re_c\gtrsim O(1)$, the inertial screening length is of order the channel width or smaller and the walls lie in the Oseen region. Fluid inertial effects now constitute a singular perturbation, and one must solve the linearized Navier-Stokes equations with the solution being a function of $Re_c$\,($\S$\ref{sec:Ch5Reclarge} below).

\section{The inertial lift velocity for $Re_c\ll1$}\label{sec:Ch5Recsmall}

In this section, we revisit the inertial lift calculation for $Re_c \ll 1$, a problem first addressed by \cite{holeal1974} and \cite{vasseur1976}. A generalized reciprocal theorem formulation yields the following expression for the inertial lift velocity \citep{anandJeffAvgd2022}:
\begin{align} 
V_p&=-Re_p\int\bm{u}^{(2)}\cdot\left(\bm{u}^{(1)'}\cdot\bm{\nabla u}^{(1)'}+\bm{u}^{(1)'}\cdot\nabla \bm{u}^\infty+\bm{u}^\infty\cdot\bm{\nabla u}^{(1)'}\right)\,dV. \label{eq:Ch5RTliftvel}
\end{align}
In (\ref{eq:Ch5RTliftvel}), problem 1 denotes the one described in $\S$\ref{sec:Ch5formulation} viz.\ a neutrally buoyant sphere suspended in wall-bounded plane Couette flow for small but finite $Re_p$, and accordingly, the disturbance fields $(\bm{u}',p')$ defined in $\S$\ref{sec:Ch5formulation} are denoted as $(\bm{u}^{(1)'},p^{(1)'})$ only in this section. Problem 2 denotes a simpler test problem, with $(\bm{u}^{(2)},p^{(2)})$ corresponding to the Stokesian translation of a torque-free sphere between plane parallel walls, under the action of a unit force in the wall-normal direction. The test problem is governed by: 
\begin{subequations}
\begin{align}
\nabla^2 \bm{u}^{(2)}-\bm{\nabla} p^{(2)} &= 0 \label{eq:Ch5NStest},\\
\bm{\nabla}\cdot\bm{u}^{(2)}&=0,
\end{align}
\end{subequations}
where $\bm{u}^{(2)}$ satisfies:
\begin{subequations}
	\begin{align} 
	\bm{u}^{(2)}&=\bm{U}_p^{(2)} \text{ for } \bm{r}\in S_p,\\
	\bm{u}^{(2)} &\rightarrow 0 \text{ for } r_1,r_3 \rightarrow \infty,\\
	\bm{u}^{(2)}&=0 \text{ for } r_2=-s\lambda^{-1}, (1-s)\lambda^{-1}.
	\end{align} \label{eq:Ch5BCtest}
\end{subequations}

To $O(Re_p)$, $\bm{u}^{(1)'}$ in \eqref{eq:Ch5RTliftvel} may be replaced by its Stokesian approximation, $\bm{u}_s^{(1)}$, the resulting volume integral being convergent. To infer the dominant scales contributing to this integral, it suffices to use estimates of the disturbance fields pertaining to the interval $1 \ll r \ll \lambda^{-1}$. Thus, using $\bm{u}^\infty\sim r$, $\bm{u}^{(2)}\sim 1/r$, $\bm{u}^{(1)'}\approx \bm{u}_s^{(1)} \sim 1/r^2$ yields $\bm{u}^{(2)}\cdot(\bm{u}_s^{(1)}\cdot\bm{\nabla u}^\infty+\,\,\bm{u}^\infty\cdot\bm{\nabla u}_s^{(1)})\sim 1/r^3$ and  $\bm{u}^{(2)}\cdot(\bm{u}_s^{(1)}\cdot\bm{\nabla}\bm{u}_s^{(1)})\sim 1/r^6$, respectively, for the linear and nonlinear components of the integrand. Since $dV \sim r^2 dr$, the dominant contributions to the integral, due to the nonlinear inertial terms, arise from length scales of $O(a)$. In contrast, the linearized inertial terms appear to lead to a conditionally convergent integral, implying that the dominant contribution must arise from scales intermediate between $a$ and $H$, with logarithmically smaller contributions from $r \sim O(a)$ and $O(H)$. Now, on account of the inversion symmetry of the plane Couette profile, the inertial lift owes its origin entirely to the asymmetry of the sphere location with respect to the walls - a neutrally buoyant sphere in unbounded simple shear flow experiences zero lift regardless of $Re_p$. Hence, the contributions from $r \ll O(H)$, which must involve unbounded-domain expressions for the disturbance fields at leading order, are identically zero, and the dominant contribution due to the linearized inertial terms arises from scales of $O(H)$. Further, the contribution of the linearized inertial terms may be shown to be larger, by a factor of $O(\lambda^{-1})$, than that of the nonlinear terms \citep{anandJeffAvgd2022}. This implies that, for purposes of calculating the integral in \eqref{eq:Ch5RTliftvel}, one may replace the neutrally buoyant sphere in Problem 1 by a stresslet, and the one in the test problem by a Stokeslet oriented perpendicular to the walls, with the $O(Re_p)$ lift velocity given by the following simplified integral:
\begin{align}
V_p &=-Re_p\int_{V}\bm{u}_\text{st}\cdot\left(\bm{u}_\text{str}\cdot\bm{\nabla u}^\infty+\bm{u}^\infty\cdot\bm{\nabla u}_\text{str}\right)\,dV, \label{eq:Ch5VpsmallRec}
\end{align} 
where
\begin{align}
\bm{u}_\text{st} =&\,\bm{J}\cdot\bm{1}_2\, \label{eq:stokeslet} \\
\bm{u}_\text{str} =&-\frac{20\pi\bm{E}^\infty}{3}:\frac{\partial \bm{J}}{\partial \bm{y}}.\, \label{eq:stresslet}
\end{align}
Here, $\bm{y}=s\lambda^{-1}\bm{1}_2$ is the position of the Stokeslet relative to the lower wall, and $\bm{u}_\text{st}$ and $\bm{u}_\text{str}$ are the bounded Stokeslet and stresslet velocity fields, respectively, with the expression for the second order tensor $\bm{J}$ given in Appendix \ref{sec:appA}; $\bm{E}^\infty =\frac{1}{2}({\bm 1}_1{\bm 1}_2+ {\bm 1}_2{\bm 1}_1)$ is the rate of strain tensor of the ambient Couette flow.

\begin{figure}
	\centering
	\includegraphics[width=\textwidth]{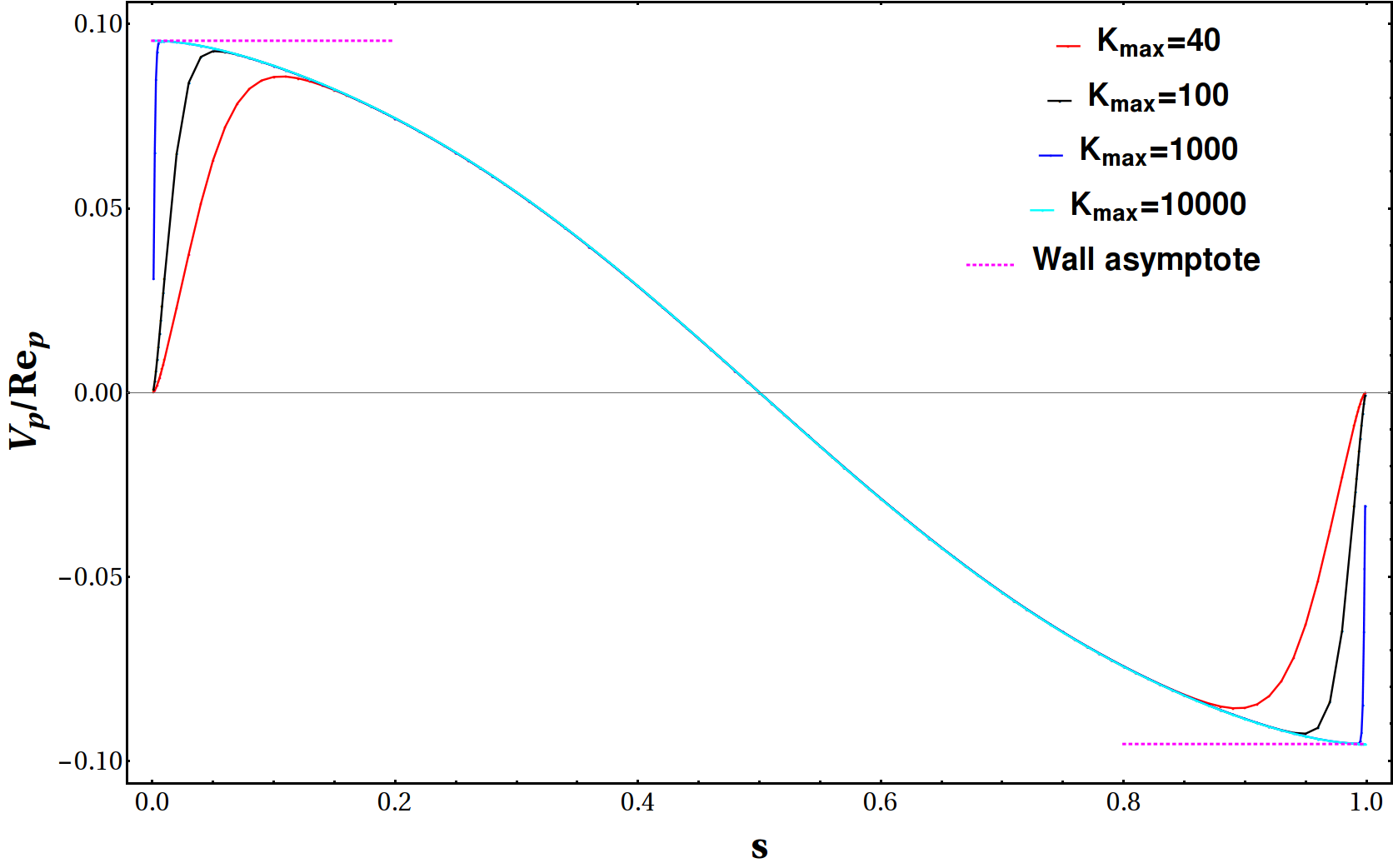}
	\caption{Lift velocity profiles for a sphere in plane Couette flow, for $Re_c\ll1$, obtained using (\ref{eq:Ch5VpExpression}) and the indicated values of $K_\text{max}$. The wall asymptotes calculated using  \eqref{eq:Ch5VpnearWall} are denoted by dotted horizontal lines.}
	\label{fig:Ch5KmaxeffectsmallRec}
\end{figure}

A more general version of \eqref{eq:Ch5VpsmallRec}, pertaining to plane Poiseuille flow, has been evaluated in \cite{anandJeffAvgd2022}. The one specific to plane Couette flow may be obtained by using $\kappa=1$, $\beta=1$ and $\gamma''=0$ in equation (87) of \cite{anandJeffAvgd2022}, and is given by:
\begin{align}
\frac{V_p}{Re_p}= -\frac{10\pi}{3} \int_0^\infty dk_\perp'' \dfrac{k_\perp''\,\,e^{-k_\perp'' (27 s+16)} I(k_\perp'',s)}{48 \pi \left(e^{2 k_\perp''}-1\right) \left[-2 e^{2 k_\perp''} \left(2 k_\perp''^2+1\right)+e^{4 k_\perp''}+1\right]^2},
\label{eq:Ch5VpExpression}
\end{align}
where the expression for $I(k_\perp'',s)$ is given in Appendix \ref{sec:appB}. The integral in \eqref{eq:Ch5VpExpression} is readily evaluated using Gauss-Legendre quadrature with a suitably large cutoff $K_\text{max}$ for the upper limit. The choice of $K_\text{max}$ is crucial close to the walls - as shown in Figure \ref{fig:Ch5KmaxeffectsmallRec}, for any finite $K_\text{max}$, $V_p$ decreases to zero in the neighborhood of the walls, the neighborhood shrinking with increasing $K_\text{max}$. For $K_\text{max} = 10000$, a numerically converged profile is obtained for $0.001\leq s\leq 0.999$.

In the neighborhood of the walls\,($s = 0,1$), the primary contribution to the integral in \eqref{eq:Ch5VpExpression} comes from $k_\perp$ of $O(s^{-1})$ or $O(1-s)^{-1}$, so the dominant length scales are of order the small separation between the sphere and the wall. This also implies that the limiting wall value obtained below may also be derived as the far-field limit of the single-wall problem \citep{cherukat1994}. Considering the wall at $s = 0$, for instance, and using a rescaled wavenumber $k_w=k_\perp'' s$, gives:  
\begin{align}
V_p^\text{wall} = \lim_{s \rightarrow 0} V_p = \frac{5 Re_p}{72} \int_0^\infty dk_w\,\,e^{-2 k_w} k_w \left(3 k_w^2 -2 k_w+3\right),
\label{eq:Ch5VpnearWall}
\end{align}
which may be evaluated analytically and gives $55Re_p/576$; the $O(s)$ correction to this asymptote involves length scales of $O(H)$. It is important to note here that the actual lift velocity must go to zero at the wall on account of the diverging lubrication resistance. Thus, the finite wall value obtained above must be interpreted as corresponding to the intermediate asymptotic interval $\lambda\ll s\ll1$\,(for small $Re_c$); this aspect is implicit in the connection to the single-wall problem mentioned above.

\begin{figure}
	\centering
	\includegraphics[width=\textwidth]{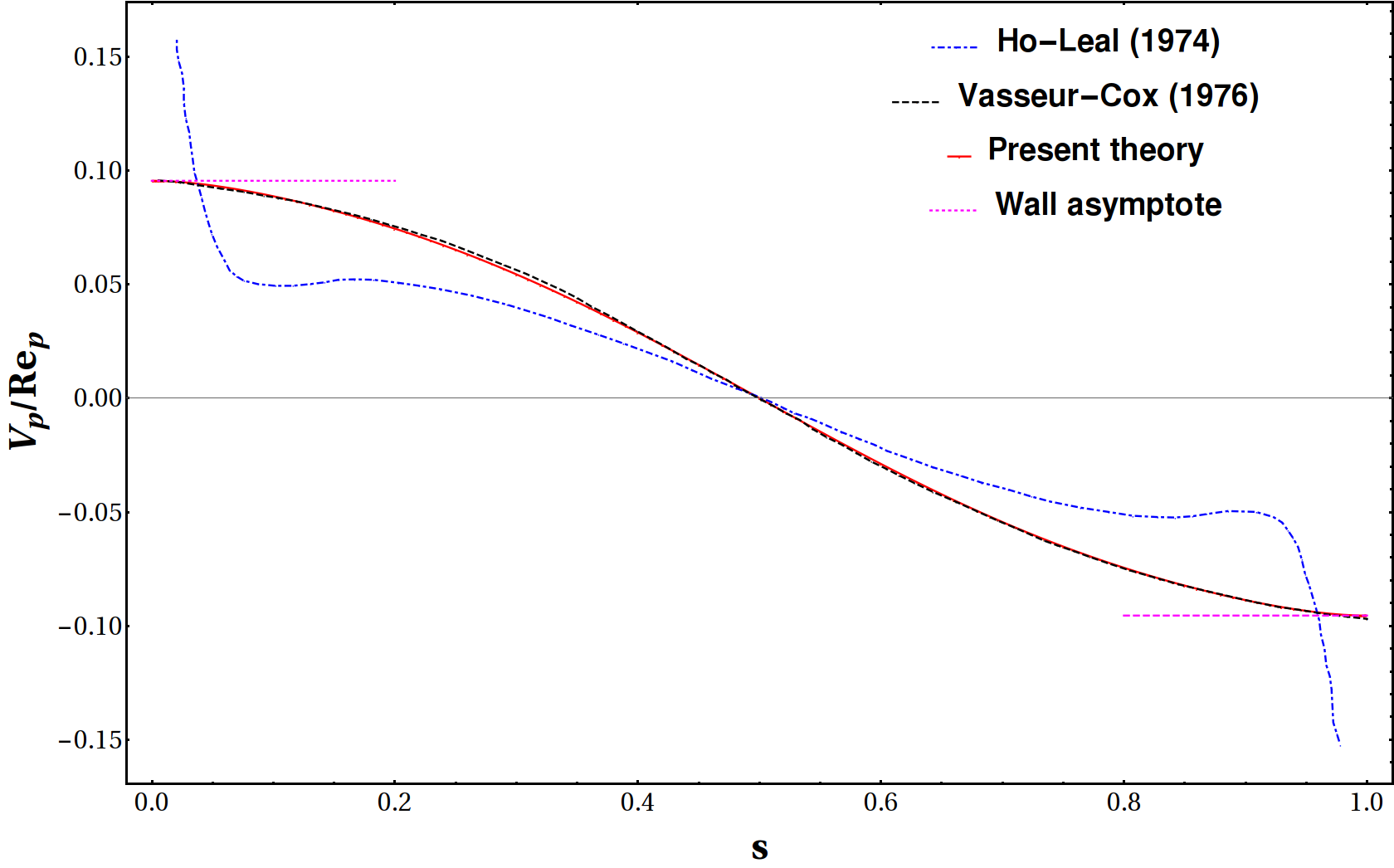}
	\caption{Lift velocity profile for a sphere in plane Couette flow for $Re_c\ll1$ compared against the data obtained from digitizing figures 2 and 3 in \cite{holeal1974} and \cite{vasseur1976}, respectively; the pair of wall asymptotes appear as horizontal dashed lines.}
	\label{fig:Ch5smallRecComparison}
\end{figure}

Figure \ref{fig:Ch5smallRecComparison} compares the lift velocity profile obtained from \eqref{eq:Ch5VpExpression}, with  $K_\text{max} = 10000$, against profiles extracted from \cite{holeal1974} and \cite{vasseur1976}. Agreement with the \cite{holeal1974} profile is poor in general, and especially near the walls. In contrast, a near-exact match is obtained with the \cite{vasseur1976} profile throughout the channel. The wall asymptote above is also shown, and is consistent with the limiting values of our profile, and the one in \cite{vasseur1976}; note that the aforementioned wall asymptote was also mentioned in \cite{vasseur1976}, albeit without any explanation. Finally, as evident from the profiles shown, the only equilibrium is the stable one at the channel centerline.

\section{The inertial lift velocity for $Re_c\gtrsim O(1)$} \label{sec:Ch5Reclarge}

For a general shearing flow, the primary contribution to the inertial lift for $Re_c \gtrsim O(1)$ comes from scales of $O(H Re_c^{-1/2})$. For plane Couette flow, however, as already pointed out, the lift arises solely due to the asymmetric interaction of the sphere with the boundaries, and the relevant scales are $O(H)$ regardless of $Re_c$. Now, for $Re_p\ll1$, either $H$ or $HRe_c^{-\frac{1}{2}}$ is much larger than $O(a)$, and determining the lift for $Re_c \gtrsim O(1)$, at leading order, requires solving the linearized Navier-Stokes equations, with the neutrally buoyant sphere treated as the same point-singularity\,(a stresslet) as in $\S$\ref{sec:Ch5Recsmall}. It is appropriate to use $H$ as the relevant length scale by defining $\bm{r}=\lambda^{-1} \bm{R}$, with the rescalings $\bm{u}'=\lambda^2 \bm{U}$ and $p'=\lambda^3 P$ for the velocity and pressure fields based on the Stokesian rates of decay for a stresslet. From (\ref{eq:Ch5NS2}a,b), $\bm{U}$ and $P$ are seen to satisfy the following equations:  
\begin{subequations}
	\begin{align}
	\nabla^2 \bm{U}-\bm{\nabla}P &=Re_c U_2\bm{1}_1+Re_c R_2\frac{\partial \bm{U}}{\partial R_1}-\frac{20\pi}{3}\bm{E}^\infty\cdot\bm{\nabla}\delta(\bm{R}),\\
	\bm{\nabla}\cdot \bm{U}&=0,
	\end{align} \label{eq:Ch5NSHinchOuterEqn}
\end{subequations}
with:
\begin{subequations}
	\begin{align}
	\bm{U} &\sim -\frac{5}{2}\frac{\bm{R}(\bm{E}^\infty:\bm{R}\bm{R})}{R^5} \text{ for } \bm{R}\rightarrow 0, \label{matching:req}\\
	\bm{U}&=0 \text{ at } R_2=-s, 1-s,  \label{eq:Ch5NSHinchOuterBC}
	\end{align}
\end{subequations}
The neutrally buoyant sphere appears as a stresslet forcing, this being the final term on the RHS of (\ref{eq:Ch5NSHinchOuterEqn}a) with $E^\infty_{ij}$ the rate of strain tensor of the ambient plane Couette flow as before. (\ref{matching:req}) is the requirement of matching with the Stokesian field in the inner region, and defined earlier in (\ref{eq:stresslet}). Following \cite{schonberghinch1989}, we define a partial Fourier transform:
\begin{align}
\hat{f}(k_1,R_2,k_3)=\int_{-\infty}^\infty\int_{-\infty}^\infty \,f(\bm{R})\,e^{\iota (k_1 R_1+ k_3 R_3)} dR_1 dR_3,
\end{align}
which yields the following coupled ordinary differential equations for the transformed pressure and normal velocity fields:
\begin{subequations} 
	\begin{align} 
	\frac{d^2\hat{P}}{dR_2^2}-k_\perp^2\hat{P}&=2\iota k_1 Re_c\hat{U}_2,\\
	\frac{d^2\hat{U}_2}{dR_2^2}-k_\perp^2\hat{U}_2&=\frac{d\hat{P}}{dR_2}-\iota k_1 Re_c R_2\hat{U}_2,
	\end{align} \label{eq:Ch5HinchODEs}
\end{subequations}
where $k_\perp^2=k_1^2+k_3^2$, with the matching condition 
	\begin{align}
	\hat{U}_2 &\sim-\frac{5\pi\iota k_1 |R_2|\,\,e^{-k_\perp |R_2|}}{3}  \text{ for } k_1,k_3\to\infty \text{ and } R_2\rightarrow 0,
	\label{eq:Ch5HinchMatchingCondns}
	\end{align} 
and the wall boundary conditions
	\begin{align}
	\hat{U}_2=\frac{d\hat{U}_2}{dR_2}&=0\,\, \text{ at }\,\, R_2=-s, 1-s.
	\label{eq:Ch5HinchBCs}
	\end{align} 
The boundary conditions above must be supplemented by the following jump conditions which relate the limiting values of the transformed fields above and below the location\,($R_2 = 0$) of the stresslet forcing \citep{anandJeffAvgd2022}:
\begin{subequations}
	\begin{align}
	\hat{P}^+(k_1,0,k_3)-\hat{P}^-(k_1,0,k_3) &=-\frac{20\pi\iota k_1}{3},\\
	\frac{d\hat{P}^+}{dR_2}(k_1,0,k_3)&=\frac{d\hat{P}^-}{dR_2}(k_1,0,k_3),\\
	\hat{U}_2^+(k_1,0,k_3)&=\hat{U}_2^-(k_1,0,k_3),\\
	\frac{d\hat{U}_2^+}{dR_2}(k_1,0,k_3)-\frac{d\hat{U}_2^-}{dR_2}(k_1,0,k_3) &=-\frac{10\pi\iota k_1}{3}.
	\end{align} \label{eq:Ch5JumpCondns}
\end{subequations}
Here, the superscripts `$-$' and `+' pertain to the intervals $-s\leq R_2<0$ and $0<R_2\leq(1-s)$, respectively. In the matching region, $\bm{U}$ must reduce to the sum of a singular stresslet contribution at leading order and a uniform flow in the transverse direction that arises due to inertia. Since the sphere is force-free, it must be convected by the latter uniform flow. The lift velocity may be determined via an inverse Fourier transform after removing\,(for numerical convenience) the normal component of the stresslet contribution:
\begin{align}
V_p&=\frac{Re_p}{4\pi^2 Re_c}\,\,\Re\left\{\int_{-\infty}^\infty\int_{-\infty}^\infty \,\hat{U}_2^\pm(k_1,0,k_3)\,dk_1\, dk_3\right\},
\label{eq:Ch5VpHinchFinal}
\end{align}
where $\Re\{.\}$ denotes the real part of a complex-valued function, and automatically achieves the removal of the purely imaginary stresslet contribution\,(see (\ref{eq:Ch5HinchMatchingCondns})). As indicated in (\ref{eq:Ch5VpHinchFinal}), one may use either $\hat{U}_2^-$ or $\hat{U}_2^+$ on account of continuity; see (\ref{eq:Ch5JumpCondns}c). The ODEs (\ref{eq:Ch5HinchODEs}a,b) along with the boundary conditions \eqref{eq:Ch5HinchBCs} and jump conditions (\ref{eq:Ch5JumpCondns}a-d) are solved using the shooting method described in Appendix A of \cite{schmid2002stability}. After computing $\hat{U}_2$, the inverse Fourier transform \eqref{eq:Ch5VpHinchFinal} is evaluated using Gauss-Legendre quadrature in a truncated domain that is a circle of a large but finite radius $K_m$. Convergence is accelerated by supplementing the numerical integral with a large-$k_\perp$ asymptote calculated along lines outlined in \cite{hogg1994}, and given by:
\begin{align}
V_p^\text{far field}= \frac{5Re_p}{576} e^{-2 \zeta}\left(12\zeta^3+10 \zeta^2+22\zeta+11\right),
\label{eq:Vplargek}
\end{align}
where $\zeta=K_m Re_c^{1/2}s$. 

\section{Results} \label{sec:Ch5Results}
\begin{figure}
	\centering
	\includegraphics[width=\textwidth]{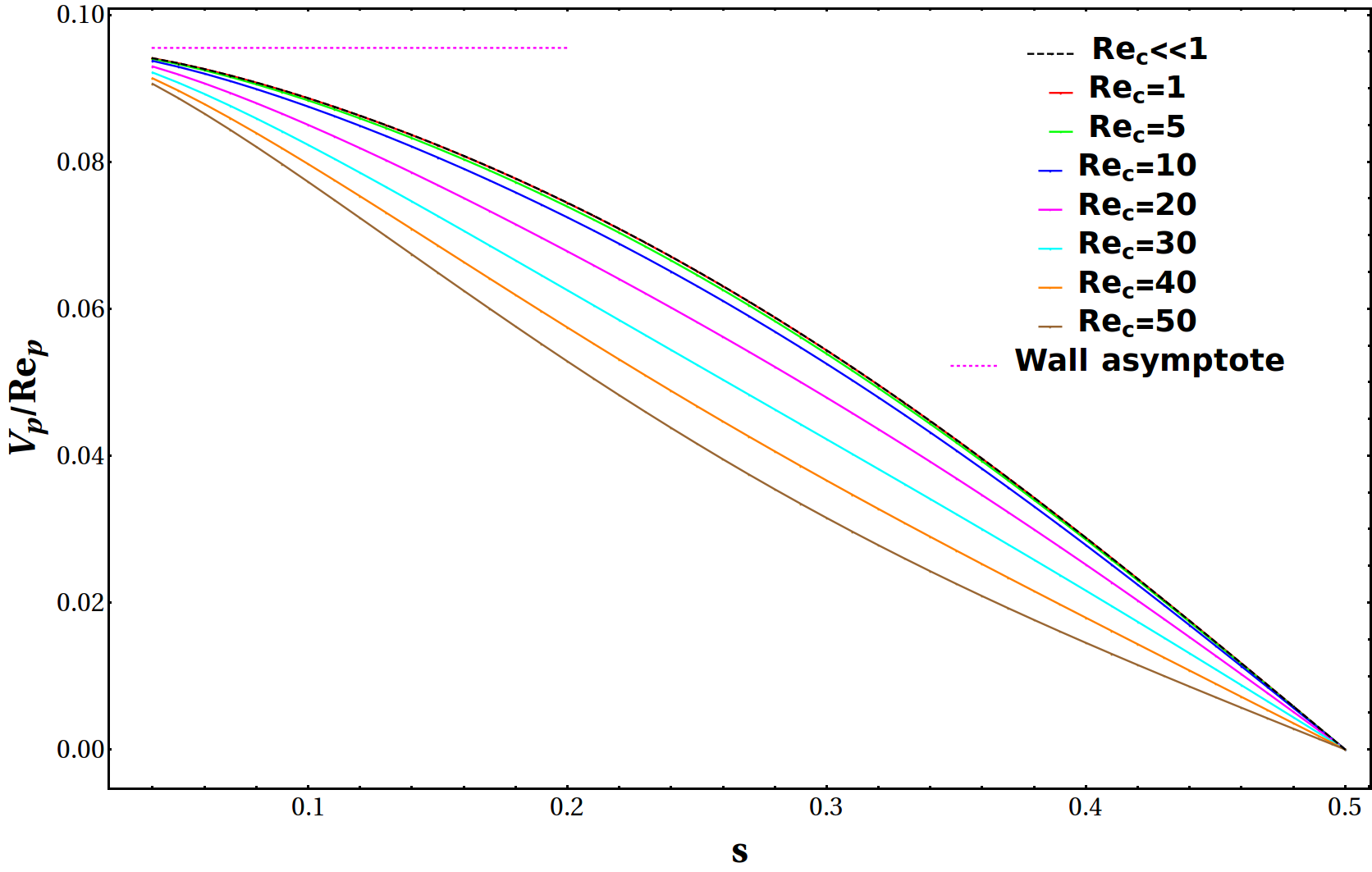}
	\caption{Lift velocity profiles for a sphere in plane Couette flow for $Re_c\gtrsim O(1)$, compared to the limiting profile for $Re_c\ll1$\,(defined by (\ref{eq:Ch5VpExpression}) in $\S$\ref{sec:Ch5Recsmall}).}
	\label{fig:Ch5orderunityRecComparison}
\end{figure}

We first validate our calculation by comparing the lift velocity profiles, for small $Re_c$, computed using the shooting method, against the one calculated in $\S$\ref{sec:Ch5Recsmall}, for $Re_c \rightarrow 0$, using a reciprocal theorem formulation; note that all profiles from hereon are only plotted over half the channel domain on account of their anti-symmetry about the centerline. As evident from Figure \ref{fig:Ch5orderunityRecComparison}, the small-$Re_c$ limiting form given by (\ref{eq:Ch5VpExpression}) remains an excellent approximation up until $Re_c\approx5$. Further, the wall value obtained in \eqref{eq:Ch5VpnearWall}, again within a small-$Re_c$ framework, remains valid for $Re_c\gtrsim O(1)$, implying that the dominant contribution to the near-wall lift comes from scales of order the small sphere-wall separation 
regardless of $Re_c$. Interestingly, Figure \ref{fig:Ch5orderunityRecComparison} shows that the inertial lift profile changes qualitatively with increasing $Re_c$. The profiles for $Re_c<40$ are concave-downward, while those for $Re_c=40$ and $50$ exhibit a concave-upward form. This change in curvature is accompanied by a flattening of the profile near the centerline, leading to a progressive decrease in the stability of the centerline equilibrium.

\begin{figure}
	\centering
	\begin{subfigure}[b]{0.49\textwidth}
	\includegraphics[width=\textwidth]{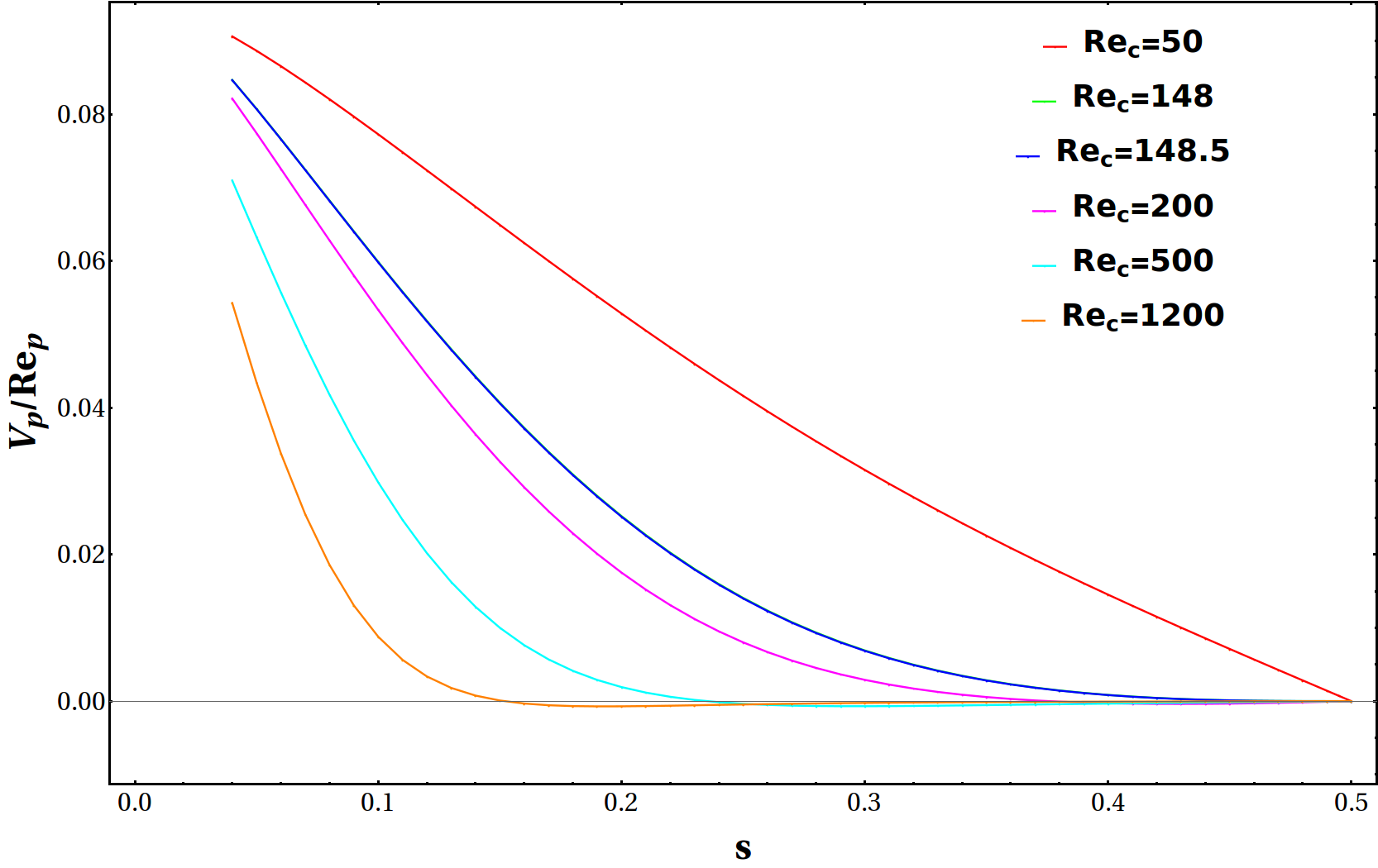}
	\caption{}
	\end{subfigure}
	\hfill
	\begin{subfigure}[b]{0.49\textwidth}
	\includegraphics[width=\textwidth]{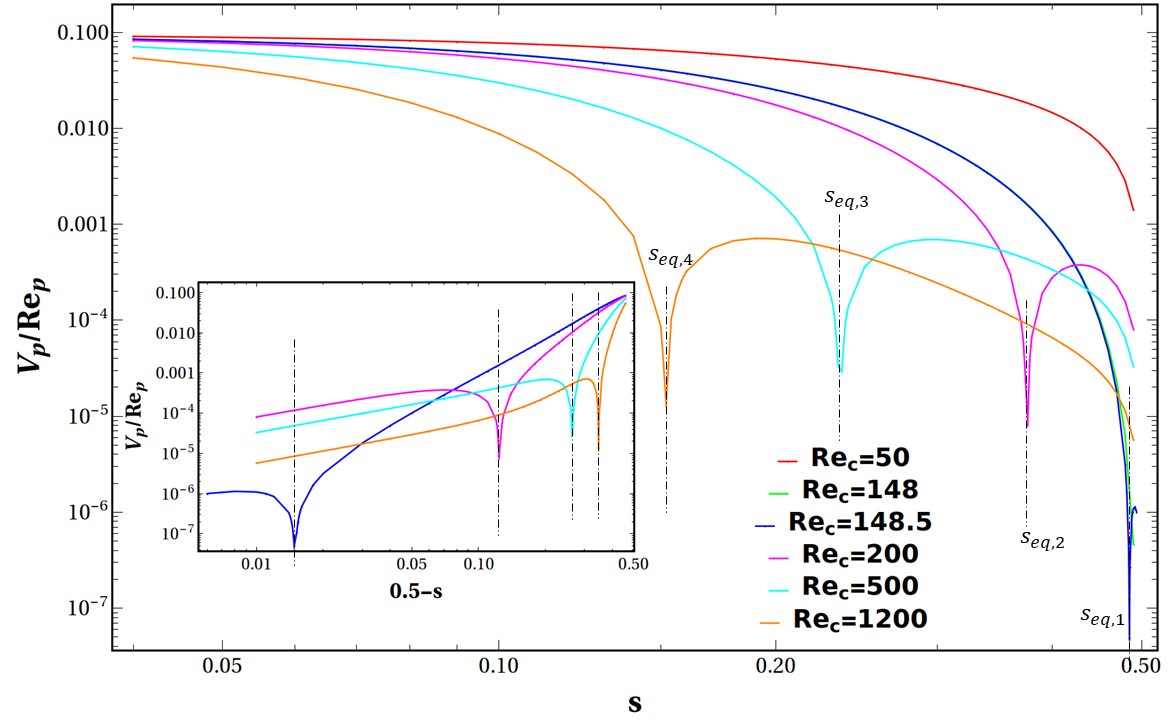}
	\caption{}
	\end{subfigure}
	\caption{Lift velocity profiles for a sphere in plane Couette flow for $Re_c\geq50$ on a) linear, and b) logarithmic scales. The linear profiles highlight the decrease in the stability of the centerline equilibrium with increasing $Re_c$; the equilibrium turns unstable for $Re_c\gtrsim 148.5$. The log-log profiles highlight the emergence of an off-center equilibrium\,($s_{eq}$), for $Re_c\geq148.5$, which then shifts towards the lower wall with increasing $Re_c$, as evident from the sequence: $s_{eq,1}=0.485$, $s_{eq,2}=0.375$, $s_{eq,3}=0.234$ and $s_{eq,4}=0.152$.}
	\label{fig:Ch5highRecplots}
\end{figure}

To explore the stability of the centerline equilibrium in more detail, we plot lift velocity profiles for higher $Re_c$'s in Figures \ref{fig:Ch5highRecplots}a and b. As $Re_c$ increases further, the aforementioned flattening becomes more pronounced, culminating in the appearance of a stable off-center equilibrium\,($s_{eq}$) at $Re_c\approx148$, with the original centerline equilibrium simultaneously becoming unstable\,(a second equilibrium in the other half of the channel is implied by symmetry). This validates the original discovery of \cite{fox2020} and \cite{fox2021} within the framework of a small-$Re_p$ point-particle formulation. The emergence of the new equilibrium is seen more clearly on the log-log plot in Figure \ref{fig:Ch5highRecplots}b, where the zero crossings corresponding to equilibria appear as dips to negative infinity\,(marked by vertical dashed lines). The inset in this figure, with $0.5-s$ as the abscissa, shows the wallward\,(lower wall) migration of the off-center equilibrium with $Re_c$ increasing beyond $148$.
% \begin{figure}
% 	\centering
% 	\includegraphics[width=0.49\textwidth]{Figs/sphere_Couette_flow_Rec_50_to_1200.png}
% 	\caption{Lift velocity profiles for a sphere in plane Couette flow for $Re_c\geq50$. The profiles highlight the decrease in the stability (of the centerline equilibrium) with increasing $Re_c$ upto $148$, with the profiles for $Re_c\geq148.5$ showing a transition to eventual instability.}
% 	\label{fig:Ch5highRecplots}
% \end{figure}
% \begin{figure}
% 	\centering
% 	\includegraphics[width=\textwidth]{Figs/sphere_Couette_flow_Rec_50_to_1200_loglog.png}
% 	\caption{Lift velocity profiles from \ref{fig:Ch5highRecplots} on a log-log scale for large $Re_c$'s highlighting the emergence of an off-centerline equilibrium for $Re_c\geq148.5$. The dashed vertical lines highlight the intermediate equilibria $s_{eq}$, which shifts towards the channel walls with increasing $Re_c$ as evident from the following sequence: $s_{eq,1}=0.485$, $s_{eq,2}=0.375$, $s_{eq,3}=0.234$ and $s_{eq,4}=0.152$.}
% 	\label{fig:Ch5profilesloglog}
% \end{figure}

\begin{figure}
	\includegraphics[width=\textwidth]{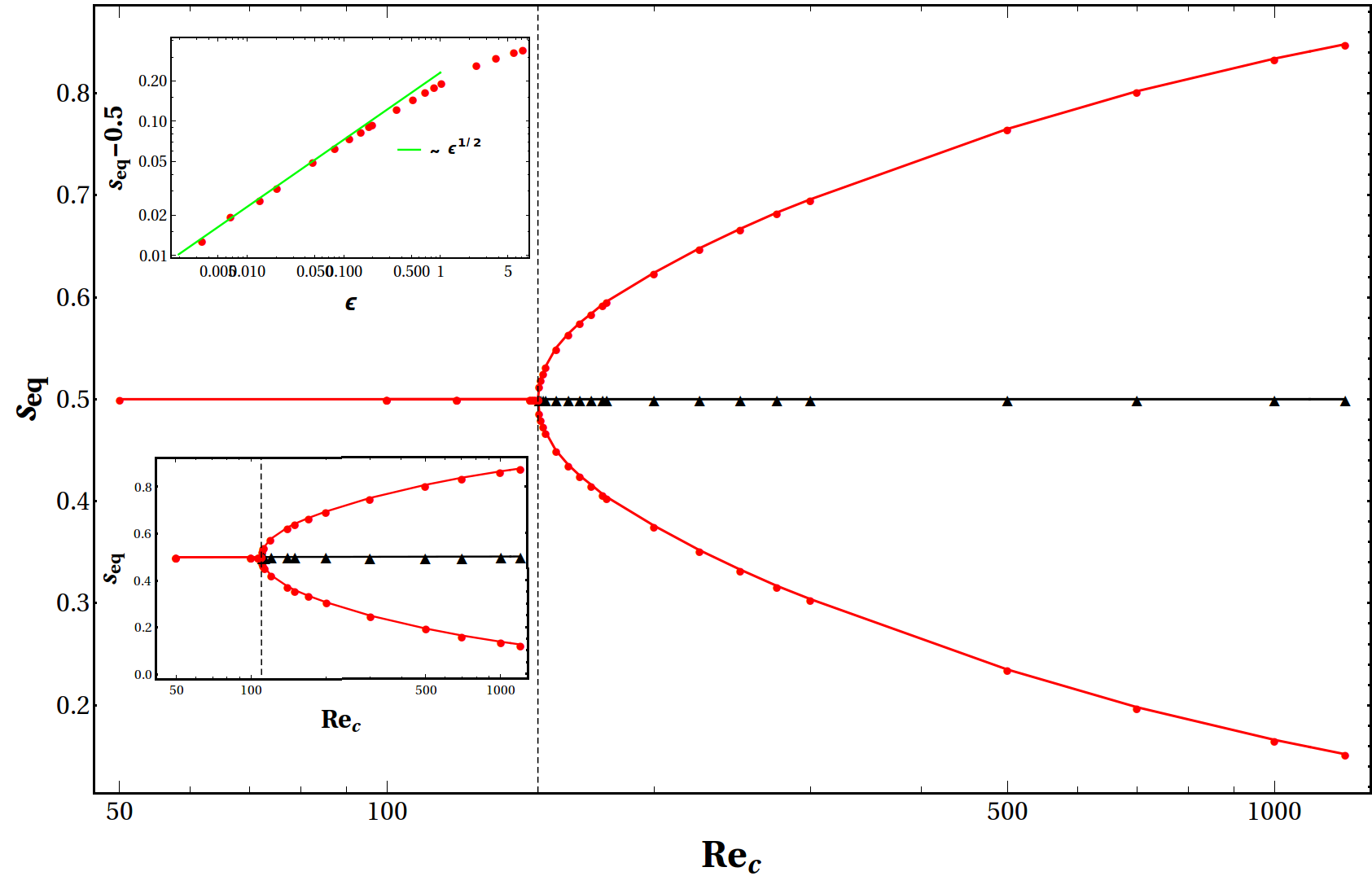}
	\caption{Equilibrium loci for a sphere and cylinder\,(lower inset) suspended freely in plane Couette flow depicting the supercritical pitchfork bifurcation; the vertical dashed lines denote $Re_c^\text{critical}$ for the two cases. The upper inset demonstrates the square-root scaling in the vicinity of the bifurcation threshold; here, $\epsilon=\Big(\dfrac{Re_c-Re_c^\text{critical}}{Re_c^\text{critical}}\Big)^{1/2}$.}
	\label{fig:Ch5equilibriumloci}
\end{figure}

The lift-force equilibria identified above are plotted as a function of $Re_c$ in Figure \ref{fig:Ch5equilibriumloci}, the resulting locus conforming to a supercritical pitchfork bifurcation; the red dots and black triangles correspond to stable and unstable equilibria, respectively. Thus, the central branch of the pitchfork corresponds to the centerline equilibrium that is stable for $Re_c \lesssim 148$, but unstable for larger $Re_c$. The two peripheral branches, consisting entirely of red dots, mark the emergence and subsequent wallward migration of the stable off-center equilibria with increasing $Re_c$. The inset on top validates the square root scaling expected in the neighborhood of the bifurcation threshold defined by $\frac{Re_c-Re_c^\text{critical}}{Re_c^\text{critical}} \ll1$\,($Re_c^\text{critical} \approx 148$ as mentioned above). The lower inset shows the analogous bifurcation for a circular cylinder with $Re_c \approx 110$ being the bifurcation threshold; the lower value of the threshold is consistent with the larger disturbance field, and the resulting stronger interaction with the walls, in two dimensions. 

While the bifurcation in Figure \ref{fig:Ch5equilibriumloci} is similar to that in figure 6 of \cite{fox2021}, there is a key  distinction that needs emphasis - the emergence of the bifurcation within a point-particle formulation, valid for small but finite $Re_p$, clearly shows that it corresponds to  a critical $Re_c$ \textbf{not} $Re_p$. Since $Re_p=\lambda^2 Re_c$, the bifurcation can occur for arbitrarily small $Re_p$, provided the confinement ratio is sufficiently small. The existence of the bifurcation in a point-particle formulation also implies that it likely owes its origin to a qualitative change in the disturbance velocity field on length scales larger than $aRe_p^{-\frac{1}{2}}$; a change that likely results in a sphere at the centerline experiencing an attractive interaction with its wall-induced images beyond $Re_c \approx 148$.

% \begin{figure}
% 	\centering
% 	\includegraphics[width=0.8\textwidth]{Figs/bifurcation_plot_cylinder_Couette_flow.png}
% 	\caption{Loci of the equilibrium locations $s_\text{eq}$ for a circular cylinder in plane Couette flow depicting the supercritical pitchfork bifurcation at $Re_c\approx110$.}
% 	\label{fig:Ch5equilibriumlociCylinder}
% \end{figure}

\section{Conclusion} \label{sec:Ch5conclusion}
In this paper, we have calculated the lift velocity of a freely rotating neutrally buoyant sphere suspended in wall-bounded plane Couette flow, in the limit $Re_p\ll1$, with $Re_c$ being arbitrary. Following \cite{holeal1974}, a generalized reciprocal theorem was used to calculate the lift velocity in the limit $Re_c\ll1$, with the resulting lift velocity profile exhibiting good agreement with the calculation of \cite{vasseur1976}, although the latter authors used a formal matched asymptotic expansions approach. For $Re_c\gtrsim O(1)$, the lift was obtained using a shooting method to solve the boundary value problem for the partial Fourier transform of the normal velocity field obtained from the linearized Navier-Stokes equations. The numerical results reveal the channel centerline to be the only (stable)\,equilibrium for $Re_c \lesssim 148$, with a supercritical pitchfork bifurcation creating a pair of stable equilibria, on either side of the centerline, for larger $Re_c$; the analogous bifurcation for a circular cylinder occurs for $Re_c \approx 110$. The $Re_p$-thresholds for these two cases are $Re^{\text{critical}}_p \approx 148\lambda^2$ and $110\lambda^2$, implying that the threshold $Re_p$ can be arbitrarily small for sufficiently small $\lambda$. The analogous lift calculation for plane Poiseuille flow\citep{schonberghinch1989,asmolov1999,anandJeffAvgd2022}, within a point-particle formulation, reveals no change in the number of equilibria with increasing $Re_c$; as first shown by \cite{schonberghinch1989}, the original pair of equilibria corresponding to the Segre-Silberberg pinch move towards the walls. It is thus interesting to note that the addition of ambient profile curvature actually simplifies the inertial migration problem! Further, our calculation, together with the emergence of an inner-equilibrium in plane\citep{anandfinitesize2022} and pipe Poiseuille flow, due to finite-size effects, points to a potentially rich equilibrium landscape for a sphere in Couette-Poiseuille flow.

\begin{figure}
	\centering
	\includegraphics[width=0.8\textwidth]{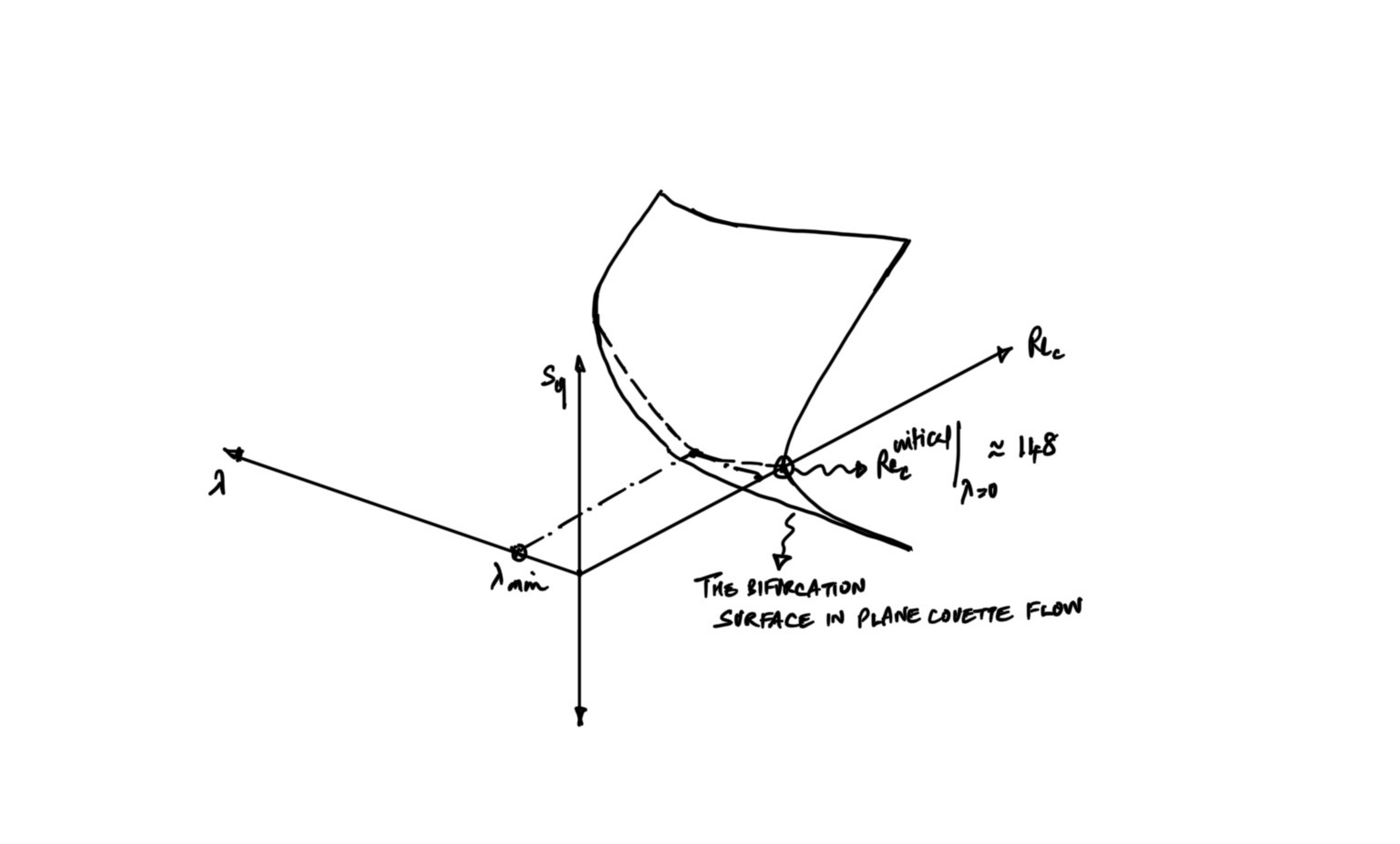}
	\caption{A tentative sketch of the bifurcation surface in the $s_{eq}-\lambda-Re_c$ space.}
	\label{fig:Ch5bifurcationSketch}
\end{figure}

The point-particle formulation here only predicts the bifurcation curve in the limit $\lambda\to0$. Thus, the locus of equilibrium positions in Figure \ref{fig:Ch5equilibriumloci} must be interpreted as the projection, onto the plane $\lambda = 0$, of a bifurcation surface in $s_{eq}-Re_c-\lambda$ space. Some idea of the nature of this surface may be obtained from the results of \cite{fox2020} and \cite{fox2021}. In the latter article, $Re_c^\text{critical}$ is found to increase from $30$ to $44$, and then to $70$, as $\lambda$ increases from $0.1$ to $0.15$ to $0.2$. Note that the $Re_c^\text{critical}$ for the smallest $\lambda\,(=0.1)$ is far smaller than the threshold value\,($\approx 148$) found here, suggesting an extremely steep decrease in the threshold as $\lambda$ increases to finite values. This pronounced sensitivity to $\lambda$ is very likely spurious, and a result of the coarse resolution, along the $Re_c$-axis, in the said simulations. This is also readily inferred from the shape of the pitchforks found in the simulations which do not conform to the square-root scaling. The less expensive simulations for a cylinder\citep{fox2020}, with a better $Re_c$ resolution\,(although not enough to recover the square-root scaling), do suggest an initial modest decrease, followed by a subsequent increase, in $Re_c^\text{critical}$ with increasing $\lambda$; the results for a cylinder also suggest a narrowing of the pitchfork with increasing $\lambda$. The sketch in Figure \ref{fig:Ch5bifurcationSketch} is a tentative depiction of the bifurcation surface based on the evidence above, and it is hoped that more comprehensive computations delineate this surface in more detail.

Extending the locus of equilibrium positions in Figure \ref{fig:Ch5equilibriumloci} to higher $Re_c$ will be limited by two factors. The first is finite-size effects which will need to be taken into account once the off-center equilibria move sufficiently close to the walls; one expects the resistance in the lubricating layer between the particle and the wall to retard and eventually arrest the wallward migration. For sufficiently small $\lambda$, the emergence of finite-size effects will likely be preceded by a transition to turbulence. This transition is subcritical, being triggered by finite-amplitude perturbations, and known to occur for $Re_c \approx 1300\!-\!1500$\citep{lundbladh1991,alfredsson1992,dauchot1995,bottin1998,lemoult2016}. Not too far above the transition threshold, one expects the off-center inertial equilibria to be smeared out into bands of a width determined by the amplitude of turbulent fluctuations. Deeper into the turbulent regime, one expects a local peak in the particle volume fraction profiles, sufficiently near the walls, on account of the underlying inertial equilibria, a feature that has indeed been observed in simulations of turbulent particle-laden plane Couette flow\,\citep{wang2017}. Weak turbulent fluctuations may also play a role in `equipartitioning' spherical particles among the pair of off-center equilibria, starting from an arbitrary initial distribution along the transverse channel coordinate; inter-particle hydrodynamic interactions may play an analogous role in the dilute limit. The analogous role of stochastic orientation fluctuations, in the context of suspension rheology, has been examined recently \citep{navaneeth2016,navaneeth2017,navaneethRapids,navaneeth_linear}.

Finally, one may comment on the implications for anisotropic particles, specifically spheroids. It has recently been shown that the Jeffery-orbit-averaged lift velocity profile for neutrally buoyant spheroids differs from that for spheres only by a proportionality factor that is a function of the Jeffery orbit constant $C$ and the spheroid aspect ratio\citep{anandJeffAvgd2022}, and therefore, the equilibrium positions for spheroids, in plane Poiseuille flow, remain identical to those for a sphere. Since the Jeffery-orbit-averaged approximation is an accurate one for spheroids with aspect ratios of order unity, the Jeffery-averaged equilibrium locus, for a neutrally buoyant spheroid in plane Couette flow, should exhibit an identical bifurcation, with emergence of stable off-centerline equilibria above $Re_c \approx 148$; note, however, that the rate of approach of a spheroid, from an arbitrary initial position, to either of the off-center equilibria will depend on the aspect ratio, in general decreasing with increasing aspect ratio.

\appendix

\section{}\label{sec:appA}
The velocity field due to the Stokeslet confined between plane parallel walls, as in the test problem, may be written as:
\begin{align}
    \bm{u}_\text{st}=\bm{J}\cdot\bm{1}_2,
\end{align}
where $\bm{J}=\bm{J}^\infty+\bm{J}^w$, $\bm{J}^\infty$ being the familiar Oseen-Burger's tensor which is given by $\bm{J}^\infty=\frac{1}{8\pi}\big(\frac{\bm{I}}{r}+\frac{\bm{rr}}{r^3}\big)$. $\bm{J}^w$ is a second order tensor that characterizes the effect of the walls, and can be obtained by solving the governing equations in the test problem and applying the no-slip condition on both the walls. This is best done via implementation of a partial Fourier transform defined by:
\begin{align}
\hat{f}(k_1,r_2,k_3)=\int_{-\infty}^\infty\int_{-\infty}^\infty \,f(\bm{r})\,e^{\iota (k_1 r_1+ k_3 r_3)} dr_1 dr_3.
\end{align}
The solution of the partially transformed equations so obtained is described in detail in \cite{anandJeffAvgd2022}\,(also see \cite{brady2010}), and yields:
\begin{align}
    \hat{\bm{J}}^w=C_{im}(\bm{k}_\perp;y_2)e^{k_\perp r_2}+D_{im}(\bm{k}_\perp;y_2)e^{-k_\perp r_2} +&\frac{1}{4k_\perp^2}\big[A_m(\bm{k}_\perp;y_2) d_i e^{-k_\perp r_2}(2k_\perp r_2+1)\nonumber\\
&+B_m(\bm{k}_\perp;y_2)\bar{d}_i e^{k_\perp r_2}(2k_\perp r_2-1)\big]. 
\end{align}
Here, $d_i=k_\perp\delta_{i2}+\iota (k_1\delta_{i1}+k_3 \delta_{i3})$ and
$\bar{d}_i=k_\perp\delta_{i2}-\iota (k_1\delta_{i1}+k_3 \delta_{i3})$ and $r_2$ is the transverse coordinate relative to the particle. The second order tensors $C_{im}$ and $D_{im}$ and the vectors $A_m$ and $B_m$ are each functions of $\bm{k}_\perp\equiv(k_1,k_3)$ and the distance of the Stokeslet $y_2$ from the lower wall, and are defined below:
\begin{align}
A_m&=\frac{Y_m \sinh (k_\perp\lambda^{-1})+Z_m k_\perp\lambda^{-1}e^{k_\perp(\lambda^{-1}-2 y_2)}}{\sinh^2 (k_\perp\lambda^{-1})-(k_\perp\lambda^{-1})^2},\\
B_m&=\frac{Y_m k_\perp\lambda^{-1} e^{-k_\perp(\lambda^{-1}-2y_2)}+ Z_m \sinh (k_\perp\lambda^{-1})}{\sinh^2 (k_\perp\lambda^{-1})-(k_\perp\lambda^{-1})^2},\\
Y_m&=-d_j(\hat{J}^\infty_{jm}|^L e^{k_\perp(\lambda^{-1}-y_2)}-\hat{J}^\infty_{jm}|^U e^{-k_\perp y_2}),\\
Z_m&=-\bar{d}_j(\hat{J}^\infty_{jm}|^L e^{-k_\perp (\lambda^{-1}-y_2)}-\hat{J}^\infty_{jm}|^U e^{k_\perp y_2}),\\
C_{im}&=\frac{F_{im}e^{-k_\perp(\lambda^{-1}-y_2)}-G_{im}e^{k_\perp y_2}}{e^{-k_\perp\lambda^{-1}}-e^{k_\perp\lambda^{-1}}},\\
D_{im}&=\frac{G_{im}e^{-k_\perp y_2}-F_{im}e^{k_\perp(\lambda^{-1}-y_2)}}{e^{-k_\perp\lambda^{-1}}-e^{k_\perp\lambda^{-1}}},\\
F_{im}&=-\hat{J}^\infty_{im}|^L-\frac{1}{4k_\perp^2}[A_m d_i e^{k_\perp y_2}(1-2k_\perp y_2)-B_m\bar{d}_i e^{-k_\perp y_2}(1+2k_\perp y_2)],\\
G_{im}&=-\hat{J}^\infty_{im}|^U-\frac{1}{4k_\perp^2}[A_m d_i e^{-k_\perp(\lambda^{-1}-y_2)}(1+2k_\perp(\lambda^{-1}-y_2))\nonumber\\
&+B_m\bar{d}_i e^{k_\perp(\lambda^{-1}-y_2)}(2k_\perp(\lambda^{-1}-y_2)-1)],
\end{align}
where the superscripts `$L$' and `$U$' denote the partially Fourier-transformed Oseen-Burger's tensor evaluated on the lower wall and upper wall, respectively. The Fourier-transformed Oseen-Burger's tensor is given by:
\begin{align}
\hat{\bm{J}}^\infty=\frac{1}{8\pi}\left(
\begin{array}{ccc}
\frac{2 e^{-k_\perp \left| r_2\right| } \pi  \left(k_3^2+k_\perp^2+k_\perp \left(k_3^2-k_\perp^2\right) \left| r_2\right| \right)}{k_\perp^3} & \frac{2 \iota e^{-k_\perp \left| r_2\right| } k_1 \pi  r_2}{k_\perp} & -\frac{2 e^{-k_\perp \left| r_2\right| } k_1 k_3 \pi (k_\perp \left| r_2\right| +1)}{k_\perp^3} \\
\frac{2 \iota e^{-k_\perp \left| r_2\right| } k_1 \pi  r_2}{k_\perp} & \frac{2 e^{-k_\perp \left| r_2\right| } \pi  (k_\perp \left| r_2\right| +1)}{k_\perp} & \frac{2 \iota e^{-k_\perp \left| r_2\right| } k_3 \pi  r_2}{k_\perp} \\
-\frac{2 e^{-k_\perp \left| r_2\right| } k_1 k_3 \pi  (k_\perp \left| r_2\right| +1)}{k_\perp^3} & \frac{2 \iota e^{-k_\perp \left| r_2\right| } k_3 \pi r_2}{k_\perp} & -\frac{2 e^{-k_\perp \left| r_2\right| } \pi  \left(k_\perp \left| \text{r2}\right|  k_3^2+k_3^2-2 k_\perp^2\right)}{k_\perp^3} \\
\end{array}
\right).
\label{eq:Ch5Jhatinfinity}
\end{align}

\section{}\label{sec:appB}
The lift velocity for a sphere in plane Couette flow for $Re_c\ll1$ is given by \eqref{eq:Ch5VpExpression}:
\begin{align}
\frac{V_p}{Re_p}= -\frac{10\pi}{3} \int_0^\infty dk_\perp'' \dfrac{k_\perp''\,\,e^{-k_\perp'' (27 s+16)} I(k_\perp'',s)}{48 \pi \left(e^{2 k_\perp''}-1\right) \left[-2 e^{2 k_\perp''} \left(2 k_\perp''^2+1\right)+e^{4 k_\perp''}+1\right]^2},
\end{align}
where $I(k_\perp'',s)$ is defined as:
\begin{align}
I(k_\perp'',s)&=-e^{k_\perp'' (25 s+18)}(s-1)^2\left[3 k_\perp''^2 (s-1)^2-2 k_\perp'' (s-1)+3\right]+e^{k_\perp'' (29 s+24)}(s-1)^2\big[3 k_\perp''^2 (s-1)^2\nonumber\\
&+2 k_\perp'' (s-1)+3\big]-2 (2 s-1) e^{3 k_\perp'' (9 s+8)} \left[6 k_\perp''^3 (s-1) s-4 k_\perp''^2 (s-1) s-3\right]\nonumber\\
&-2 (2 s-1) e^{9 k_\perp'' (3 s+2)} \left[6 k_\perp''^3 (s-1) s+4 k_\perp''^2 (s-1) s+3\right]-s^2 e^{k_\perp'' (25 s+26)} \left(3 k_\perp''^2 s^2-2 k_\perp'' s+3\right)\nonumber\\
&+s^2 e^{k_\perp'' (29 s+16)} \left(3 k_\perp''^2 s^2+2 k_\perp'' s+3\right)-2 e^{k_\perp'' (27 s+20)} \big[8 k_\perp''^4 s \big(2 s^2-3 s+1\big)\nonumber\\
&-6 k_\perp''^3 s \left(2 s^2-3 s+1\right)-12 k_\perp''^2 \left(2 s^3-3 s^2+3 s-1\right)-18 s+9\big]\nonumber\\
&+2 e^{k_\perp'' (27 s+22)} \big[8 k_\perp''^4 s \left(2 s^2-3 s+1\right)+6 k_\perp''^3 s \left(2 s^2-3 s+1\right)-12 k_\perp''^2 \left(2 s^3-3 s^2+3 s-1\right)\nonumber\\
&-18 s+9\big]+e^{5 k_\perp'' (5 s+4)} \big[12 k_\perp''^4 (s-1)^2 s^2+4 k_\perp''^3 (s-1)^2 (4 s-1)\nonumber\\
&+3 k_\perp''^2 \left(4 s^4-12 s^3+14 s^2-12 s+5\right)-2 k_\perp'' \left(4 s^3-9 s^2+9 s-3\right)+3 \left(4 s^2-6 s+3\right)\big]\nonumber\\
&-e^{k_\perp'' (29 s+22)} \big[12 k_\perp''^4 (s-1)^2 s^2-4 k_\perp''^3 (s-1)^2 (4 s-1)+3 k_\perp''^2 \left(4 s^4-12 s^3+14 s^2-12 s+5\right)\nonumber\\
&+2 k_\perp'' \left(4 s^3-9 s^2+9 s-3\right)+3 \left(4 s^2-6 s+3\right)\big]\nonumber\\
&+e^{k_\perp'' (25 s+24)} \big[12 k_\perp''^4 (s-1)^2 s^2+4 k_\perp''^3 s^2 (4 s-3)+3 k_\perp''^2 \left(4 s^4-4 s^3+2 s^2+4 s-1\right)\nonumber\\
&+k_\perp'' \left(-8 s^3+6 s^2-6 s+2\right)+12 s^2-6 s+3\big]-e^{k_\perp'' (29 s+18)} \big[12 k_\perp''^4 (s-1)^2 s^2\nonumber\\
&-4 k_\perp''^3 s^2 (4 s-3)+3 k_\perp''^2 \left(4 s^4-4 s^3+2 s^2+4 s-1\right)+k_\perp'' \left(8 s^3-6 s^2+6 s-2\right)+12 s^2\nonumber\\
&-6 s+3\big]-e^{k_\perp'' (25 s+22)} \big[24 k_\perp''^4 (s-1)^2 s^2+4 k_\perp''^3 (2 s-1)^3+3 k_\perp''^2 \big(6 s^4-12 s^3+10 s^2\nonumber\\
&-4 s+3\big)-6 k_\perp'' \left(2 s^3-3 s^2+3 s-1\right)+9 \left(2 s^2-2 s+1\right)\big]\nonumber\\
&+e^{k_\perp'' (29 s+20)} \big[24 k_\perp''^4 (s-1)^2 s^2-4 k_\perp''^3 (2 s-1)^3+3 k_\perp''^2 \left(6 s^4-12 s^3+10 s^2-4 s+3\right)\nonumber\\
&+6 k_\perp'' \left(2 s^3-3 s^2+3 s-1\right)+9 \left(2 s^2-2 s+1\right)\big].
\end{align}

\bibliographystyle{jfm}
\bibliography{references}

\end{document}